# STUDY OF LAGP COATING ON POLYETHYLENE SEPARATOR FOR POLYSULFIDE SUPPRESSION IN THIN LI-S BATTERIES

[1]Giovanni Ceccio, [1]Jiri Vacik

[1] *Department of Neutron and Ion Methods, Nuclear Physics Institute of CAS, Hlavní 130, 250 68 Řež, Prague, Czech Republic, ceccio@ujf.cas.cz*

## Abstract

The use of a thin solid electrolyte layer on polyethylene (PE) separators has been explored as a potential strategy to mitigate the polysulfide shuttle effect in lithium-sulfur (Li-S) batteries, a phenomenon responsible for significantly reducing their lifespan. While some solid electrolytes, such as LAGP ($Li_{1+x}Al_xGe_{2-x}(PO_4)_3$), have demonstrated the ability to effectively suppress the shuttle effect, their brittle ceramic nature limits their application, particularly in flexible electronics. Coating PE separators with LAGP nanofilms offers a promising solution by combining the mechanical flexibility and thinness of polymer separators with the electrochemical advantages of LAGP, due to its NASICON-like structure and high ionic conductivity. In this study, PE separators were coated with LAGP nanofilms using the ion sputtering technique and further modified with ferroic elements such as Ni and Co. The surface morphology of the coated separators was analyzed using scanning electron microscopy (SEM), and staircase voltammetry was employed to evaluate the electrochemical performance and modifications induced by the coatings. The results demonstrate that the use of PE separators with a thin LAGP layer significantly influences polysulfide diffusion and suppresses the shuttle effect, thereby reducing unwanted processes in Li-S batteries.

## *1. Introduction*

Energy storage devices are currently a hot topic in the scientific community due to their significant impact on modern society. As renewable energy storage facilities become more efficient and widespread, as the demand for advanced energy storage devices is growing. Lithium-sulfur (Li-S) batteries, along with other innovative systems, such as sodium-ion (Na-ion) batteries, are considered promising electrochemical solutions for the future of energy storage [1]. Li-S batteries offer several advantages, including their high theoretical capacity, high energy density, and low costs. However, their extensive use is hindered by poor cyclability, mainly due to the "shuttle effect" [2]. The shuttle effect arises from the migration and deposition of polysulfides on the electrode surface, which reduces conductivity and prevents further cycling.

Various strategies have been explored to reduce or suppress the shuttle effect, aiming to decrease polysulfide diffusion and increase battery lifespan. A recent approach proposes the use of solid electrolytes with superior ionic conductivity to suppress polysulfide migration by acting as an additional physical barrier [3]. In this work, we aim to improve this strategy by combining LISICON materials, commercially available flexible polyethylene (PE) separators, and metallic coatings for Li-S applications. This experiment not only provides valuable insights into the cyclability and lifetime of Li-S batteries but also serves as a preliminary study on the effectiveness of LISICON material used in thin film form.

## *2. Methodology*

Polyethylene separators with a thickness of 12 µm (Nanografi) were coated using the ion sputtering technique at the Low Energy Ion Facility (LEIF) in the NPI infrastructure [CANAM]. LEIF is a versatile ion beam facility that provides high current (in the mA range) for sputtering high-purity targets in a vacuum environment, as well as for material irradiation/implantation at high fluences with energies ranging from 1 to 35 keV. High-quality LISICON material, specifically LAGP $Li_{1+x}Al_xGe_{2-x}(PO_4)_3$, sourced from MSE Supplies), was used as the target for the deposition of ultrathin coatings onto the PE separators. The LAGP target was placed in the high-vacuum chamber, mounted on a rotating target holder capable of accommodating up to six targets. For metallic coatings, Ni and Co targets (purity 4N, from J.K. Lesker) were added to the holder. This setup allowed for the deposition of Co and Ni on the LAGP-coated PE surface, providing insight into the effects of these metals in combination with LAGP for polysulfide suppression. A total of three sets of samples were produced for this study: PE/LAGP, PE/LAGP/Co, and PE/LAGP/Ni. These samples were fabricated to investigate the contributions of Co and Ni, along with LAGP, to the suppression of polysulfide migration in Li-S batteries.

For the sputtering of the targets, an intense argon (Ar) ion beam with an energy of 20 keV and a focused beam area of 1 cm² was directed onto the target material. The PE separators were positioned (as substrates) 100 mm away from the target surface. The entire sputtering process was conducted under a base pressure of 7.6 x $10^{-6}$ mbar. A schematic representation of the experimental setup is shown in Figure 1.

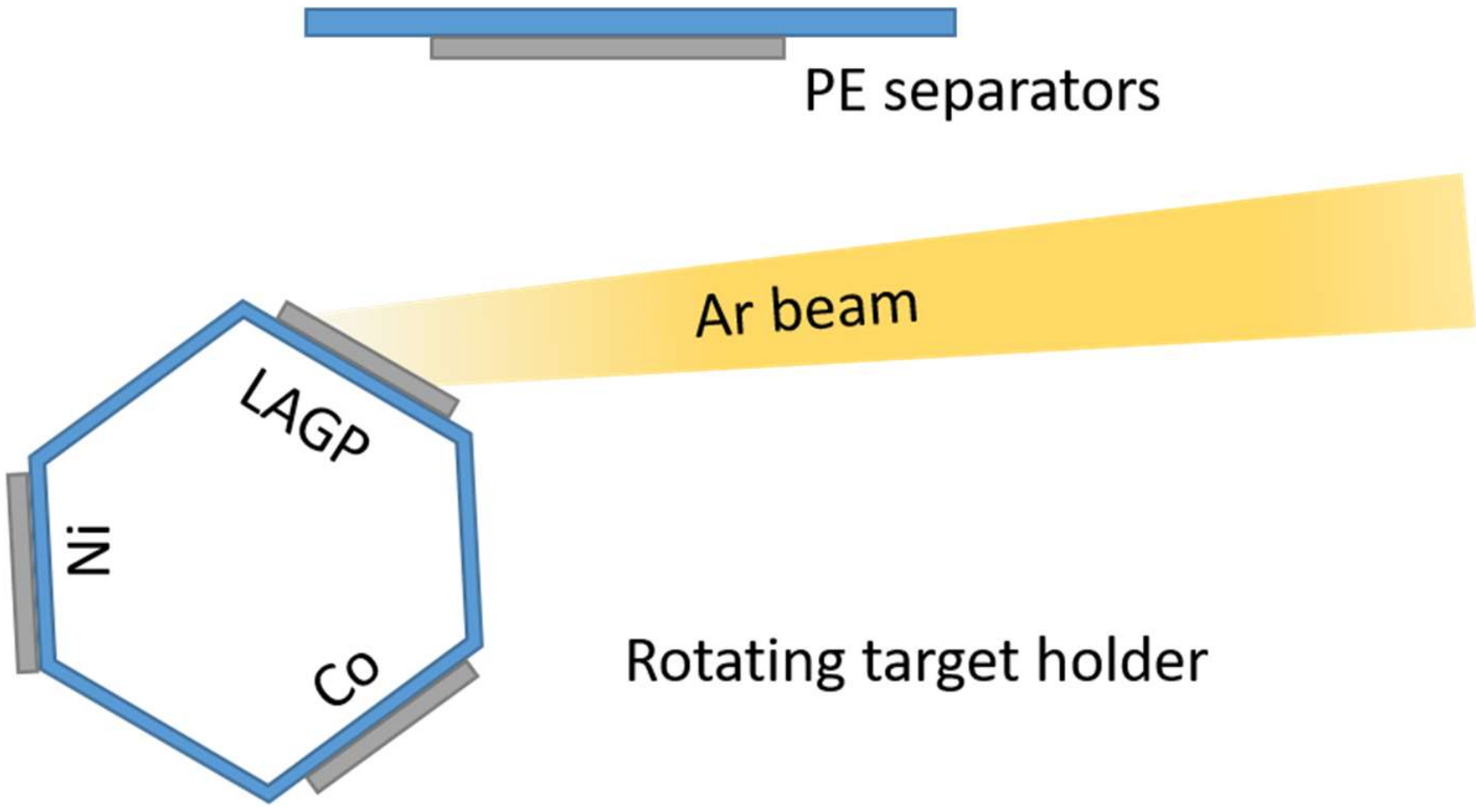

**Figure 1** Sketch of thin films deposition by ion beam sputtering.

The surface morphology of the prepared samples was analyzed using Scanning Electron Microscopy (SEM) (Hitachi TM4000 Plus). Electrochemical properties were studied using Staircase Voltammetry (SV) measurements performed with an AutoLab potentiostat (PGSTAT 204). For the CV measurements, the separators were assembled into a lab-made Teflon H-Cell. Each half-cell was filled with 7.5 mL of a 1M LiTFSI/DOL+DME (1:1, v/v) electrolyte. Sulfur (Sigma-Aldrich) was used as the cathode material, and carbon (MTI Corporation) as the anode material.

## *3. Results*

Fig. 2 show topography micrographs obtained by the SEM analysis of the prepared samples. As can be seen, the surface structure of each separators remains visible even after coating, indicating that the deposited film may not fully block or close the physical pores within the fiber-like lattice of the PE material. Additionally, the nanofilms appear homogeneous, with no noticeable micro-agglomeration or clustering of nanoparticles on the surface.

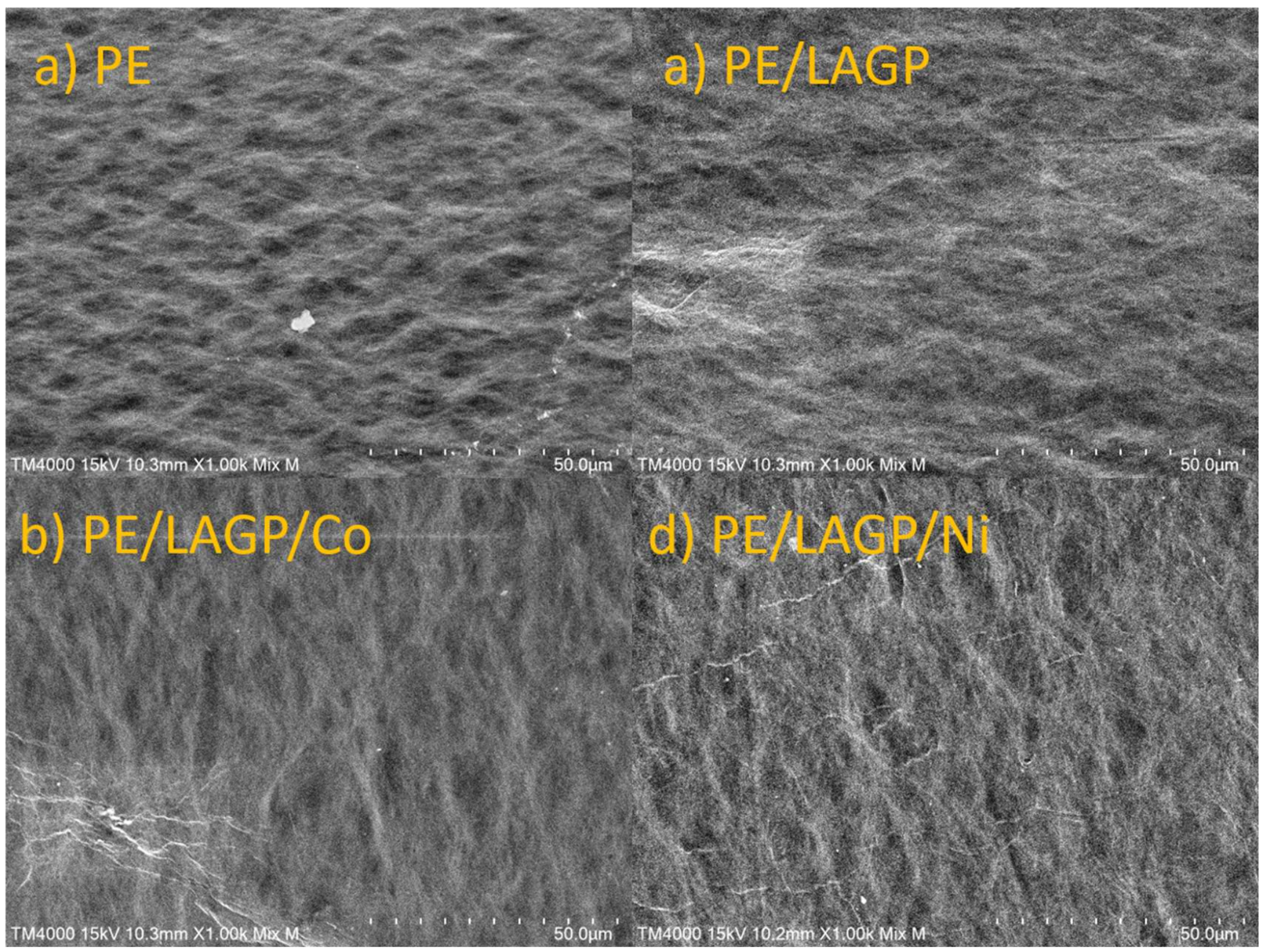

**Figure 2** SEM micrograph of the PE separator with LISICON coatings; a) shows the pristine surface of PE separator, b) the surface coated with LAGP thin film, c) shows the combination of LAGP and Co bi-layer on the surfaces d) shows the bi-layer of LAGP with Ni.

Staircase voltammetry (SV) was performed on both pristine and modified PE separators to evaluate the electrochemical behavior and performance improvements induced by the thin film coatings. The plot (Fig. 3) represents a set of cycle curves, each representing the average of five cycles for each sample, providing a reliable measure of stability and performance.

Stability and Performance: The stability over time was generally good for all the coated samples, except for the Ni-deposited PE separator, which exhibited significant instability. In the plot is shown only the first cycle for this sample due to its rapid degradation after initial cycling. This suggests that the Ni coating may not offer long-term electrochemical stability, in contrast to the LAGP-based coatings.

Effect of LAGP and Co Coatings: The introduction of LAGP significantly improved the electrochemical performance of the PE separator. More importantly, the combination of LAGP with Co further reduced the slope of the cyclic voltammogram, indicating a decrease in the reaction rate at the electrodes. This is represented by the reduction of characteristic redox peaks in the voltammetry plot, which is indicative of reduced polysulfide migration. The suppression of the polysulfide shuttle effect is a critical improvement for Li-S batteries, as it enhances the cyclability and overall lifespan of the battery.

Polysulfide Migration Reduction: Further evidence of the polysulfide migration reduction is provided by the post-test analysis of the electrolyte collected from the carbon (C) electrode side. As shown in Fig. 4, the electrolyte exhibited a significant reduction in the characteristic sulfuric color (typically associated with polysulfide species), which confirms that the presence of the coatings (especially LAGP and Co) effectively suppressed the polysulfide migration during cycling.

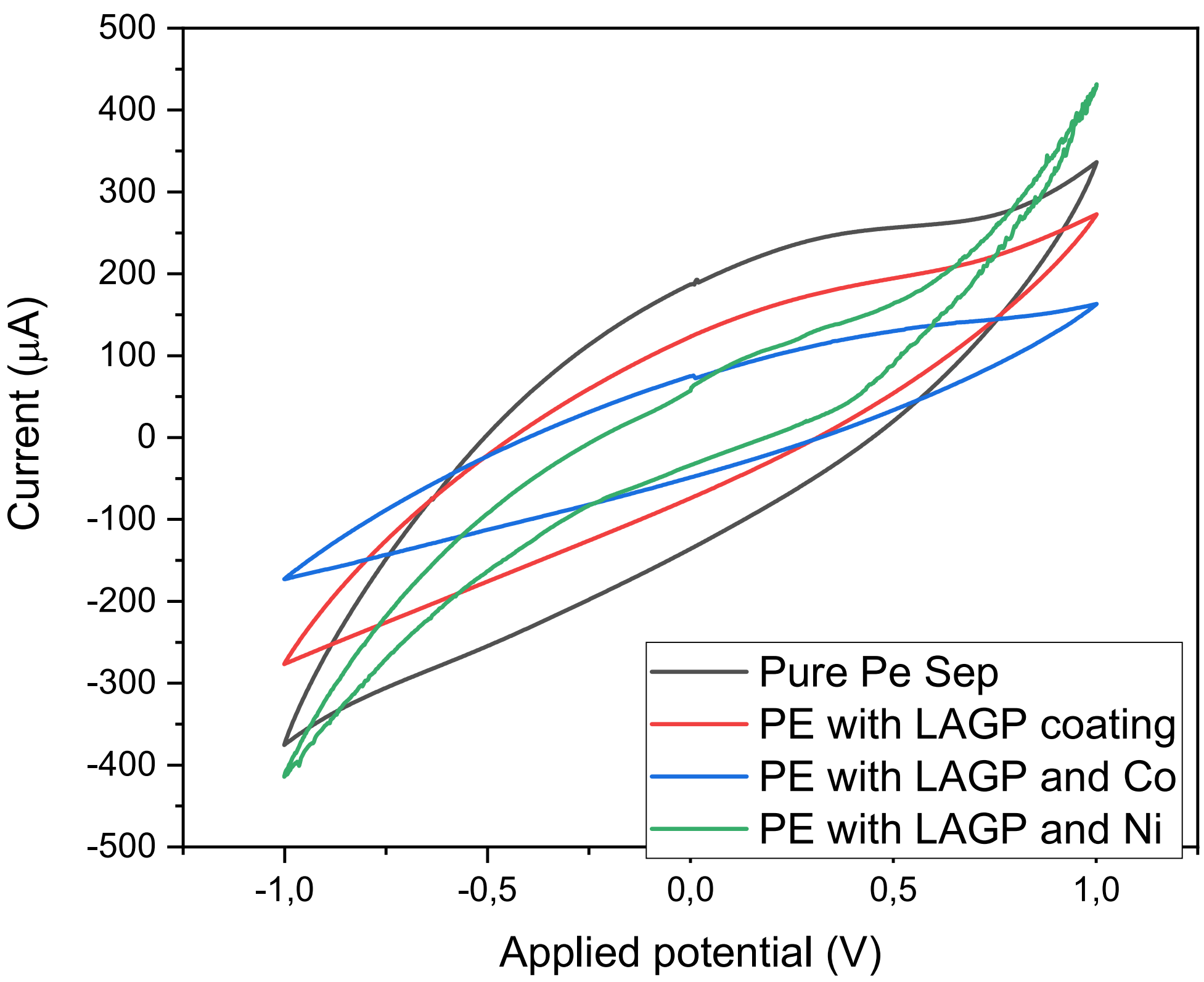

**Figure 3** Staircase voltammetry through the PE separators pristine and modified by thin films coatings.

reduced polysulfide migration. The suppression of the polysulfide shuttle effect is a critical improvement for Li-S batteries, as it enhances the cyclability and overall lifespan of the battery.

Polysulfide Migration Reduction: Further evidence of the polysulfide migration reduction is provided by the post-test analysis of the electrolyte collected from the carbon (C) electrode side. As shown in Fig. 4, the electrolyte exhibited a significant reduction in the characteristic sulfuric color (typically associated with polysulfide species), which confirms that the presence of the coatings (especially LAGP and Co) effectively suppressed the polysulfide migration during cycling.

.

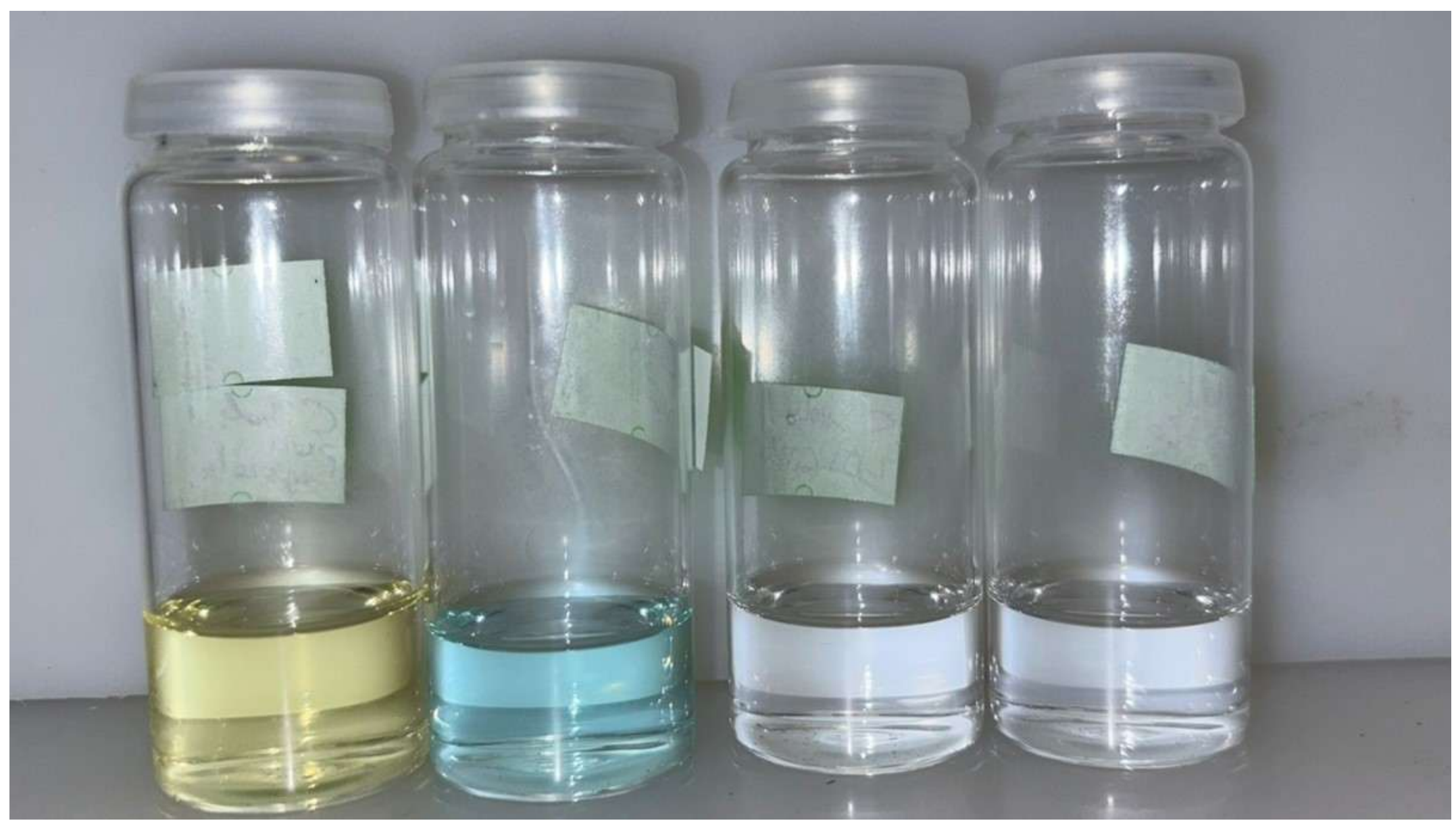

**Figure 4** The optical images of the diffusion process of polysulfides of PE separator.

## *4. conclusion*

This pilot study was designed to explore the use of low-energy ion beams for preparing solid electrolytes in the form of ultrathin films, with the aim of advancing all-solid-state thin-film batteries. The results demonstrate the successful mitigation of the shuttle effect in the presence of LISICON, particularly with ferroic elements on the surface. Staircase voltammetry data from the PE separators coated with these films show a reduction in redox peaks and a decreased slope, indicating improved electrochemical performance. Polysulfide suppression is a crucial challenge for the widespread adoption of Li-S batteries, and the findings here suggest that it can significantly enhance energy storage capabilities. Based on these results, the combination of Ni and LAGP coatings presents a promising direction for future research.

## *ACKNOWLEDGEMENTS*

*This work was supported by the Ministry of Education, Youth and Sports (MEYS) CR under the project OP JAK CZ.02.01.01/00/22_008/0004591.*